\documentclass[prl,twocolumn,showpacs,amsmath,amssymb]{revtex4}
\usepackage{graphicx}% Include figure files
\usepackage{dcolumn}% Align table columns on decimal point
\usepackage{bm}% bold math

\begin{document}
               
\title{Orientational field-dependence of low-lying excitations in mixed state
of unconventional superconductors}% 

\author{P. Miranovi\' c}
\author{N. Nakai}
\author{M. Ichioka}
\author{K. Machida}

\affiliation{Department of Physics, 
Okayama University, Okayama 700-8530, Japan}

\date{\today}

\begin{abstract}
Orientational field-dependence of the zero energy
density of states (ZEDOS) is calculated for superconductors with
the polar state (line node), axial state (point node) and 3D d-wave
state. Depending on the gap topology and relative field direction
the field dependencies of ZEDOS sensitively differ, providing us 
a useful and practical method to identify the gap topology. It is also
demonstrated that for d-wave state the field rotation in  the basal plane 
shows a sizable oscillation ($\sim$3\%) of ZEDOS. This is directly measurable
in low-T specific heat experiment in the mixed state.
\end{abstract}

\pacs{74.25.Bt, 74.25.Op}

\maketitle

Much attention has been focused on various unconventional superconductors, 
ranging from high T$_c$ cuprates, heavy Fermion materials, Boro-carbides, 
MgB$_2$, to Skutterudite PrOs$_4$Sb$_{12}$. The unconventionality is 
associated with the gap anisotropy of the orbital function in addition to 
the spin structure of the Cooper pair. It is quite important to determine 
the detailed nodal topology of the gap function; either point or line node 
and their location on the Fermi surface. The determination of these 
characteristics is expected to lead to an understanding to the pairing 
mechanism of exotic superconductors \cite{sigrist}.

There are several experimental methods to probe the gap anisotropy. One can 
basically  distinguish the line and point nodes because these give rise to 
different and distinct power law temperature (T) dependence in various 
physical quantities. As for the orientation of these nodes, there is only a 
few ways to probe it. The field-dependent thermal conductivity $\kappa(H)$ and 
the polarization dependent sound attenuation are typical ones. By measuring 
$\kappa (H,\alpha)$ for different field direction $\alpha$ one can detect 
the location of the node in 
principle because the nodal quasiparticles (QP) under $H$ with zero-energy 
transport heat current. In fact, a series of experiments by Izawa {\em et al.} 
\cite{izawa} have determined the location of nodes in several systems. 
These transport measurements are, however, inevitably involved by the 
scattering time effect and  localization effect of nodal QP, which hamper 
the determination of the nodal direction in some cases \cite{d-wave}.

Here we propose another method based on thermodynamics: The Sommerfeld 
coefficient $\gamma$ of the $T$-linear specific heat at lower $T$ is most 
fundamental physical quantity in Fermionic systems of interest. Since the 
nodal QP created around a vortex core in the mixed state sensitively reflect 
its gap structure, the angle-dependent $\gamma(H,\alpha)$ can yield 
characteristic oscillation pattern relative to the nodal position under a 
fixed $H$. Recent angle-resolved $\gamma(H,\alpha)$ measurement on 
YNi$_2$B$_2$C by Park et al \cite{park} demonstrates a fourfold oscillation 
in the basal plane whose amplitude $\sim 4\%$, nicely coinciding with 
$\kappa(H, \alpha)$ experiment by Izawa et al. \cite{izawa}. 
They agree with the nodal direction ([100]), but disagree with the topology 
(point (line) in the latter (former)). The observed oscillation 
amplitude ($\sim 4\%$) in $\gamma(H, \alpha)$ is far off the theoretical 
prediction ($\sim 30\%$) based on the so-called Doppler shift 
argument \cite{vekhter} which is qualitative in nature
(see discussion in Ref. \cite{dahm}). In this paper we 
calculate the  zero-energy density of states (ZEDOS) in the mixed state for 
various situations; its direction-dependence for the polar and axial state 
and the angular dependence for d-wave state,
in order to examine the experimental feasibility.
We develop a full three 
dimensional (3D) computation based on quasi-classical framework which is 
valid for superconductors with $k_F\xi\gg 1$ ($k_F$ Fermi wave number and 
$\xi$ the coherence length). This kind of calculations gives quantitatively
reliable results. 

Anisotropic pairing is routinely analyzed within the
separable model of pairing potential $V(\bm k,\bm k')=
V_0\Omega(\bm k)\Omega(\bm k')$. Then order parameter
takes the following form: $\Delta(\bm r,\bm k)=
\Psi(\bm r)\Omega(\bm k)$.
In the clean limit quasiclassical equations read as

\begin{equation}
\left[
2\hbar\omega_n+\hbar\bm v
\left(
\nabla+\frac{2\pi i}{\Phi_0}\bm A
\right)
\right]f=2\Psi(\bm r)\Omega(\phi,\theta)g,
\label{eil1}
\end{equation}
\begin{equation}
\left[
2\hbar\omega_n-\hbar\bm v
\left(
\nabla-\frac{2\pi i}{\Phi_0}\bm A
\right)
\right]f^\dagger=2\Psi^*(\bm r)\Omega(\phi,\theta)g.
\label{eil2}
\end{equation}
Here $\hbar\omega_n=\pi T(2n+1)$ with integer $n$ are Matsubara
frequencies, $\bm v$ is Fermi velocity, $\Phi_0$ is flux quantum,
and $f$, $f^\dagger$, $g$ are Green's functions integrated over energy
normalized so that $ff^\dagger+g^2=1$. Fermi surface is assumed to be
sphere. Order parameter $\Psi(\bm r)$ and vector-potential $\bm A(\bm r)$
are obtained selfconsistently from the following equations
\begin{equation}
\Psi(\bm r)\ln\frac{T_c}{T}=2\pi T\sum\limits_{\omega_n>0}
\left[
\frac{\Psi(\bm r)}{\hbar\omega_n}-\left<
\Omega(\phi,\theta)f
\right>
\right],
\end{equation}
\begin{equation}
\nabla\times\nabla\times\bm A(\bm r)=
-\frac{4\pi^2 \hbar N_0T}{\Phi_0}\,{\rm Im}\sum\limits_{\omega_n>0}
\left<g\bm v 
\right>.
\end{equation}
Average over Fermi surface is denoted as $\left<\ldots \right>$.
First, polar state with a line node $\Omega(\phi,\theta)=\sqrt{3}\cos\theta$
and axial state with point nodes $\Omega(\phi,\theta)=\sqrt{3/2}\sin\theta$
are analyzed. Polar and azimuthal angle refer to the
coordinate system with $z$-axis that coincide with $c$ crystal
direction. Factors $\sqrt{3}$ and $\sqrt{3/2}$ assure that average of 
$|\Omega(\phi,\theta)|^2$ over spherical Fermi surface is unity. 
We are interested in Green's function $g(\bm r,\bm v,\omega)$ 
that describe QP  excitations associated with vortices. 
The QP  density of states $N(E)$
with energy $E$ relative to Fermi level is 
defined as
\begin{equation}
\frac{N(E)}{N_0}=\frac{\left<N(E,\bm v)\right>}{N_0}=
\left<\overline{{\rm Re}\,g(\bm r,\bm v,
\omega\rightarrow 0^+-iE)}\right>,
\end{equation}
where $N_0$ is ZEDOS in the normal state. Green's function $g(\bm r)$ 
spatially averaged over vortex lattice unit cell is denoted as 
$\overline{g(\bm r)}$.
We focus on ZEDOS at low temperatures. This is because low temperature
specific heat is $C_s=\gamma(B)T=2\pi^2\hbar^2 N(E=0,B)T/3$.
Therefore the equations are solved for $T=0.1T_c$.
It is sufficient to know the Green's function only in the vortex lattice
unit cell which is divided in mesh $41\times 41$.
Once the order parameter and vector potential are obtained 
selfconsistently Eqs. (\ref{eil1}) and (\ref{eil2}) are solved 
again for $\omega\longrightarrow 0^+$. Typically we choose 
$\omega=0.001\pi T_c/\hbar$. Method of solution is extended from 
\cite{ichioka} and details will be described
elsewhere \cite{pedjanew}. Here we present the results.

\begin{figure}[h]
\includegraphics[angle=0,scale=0.45]{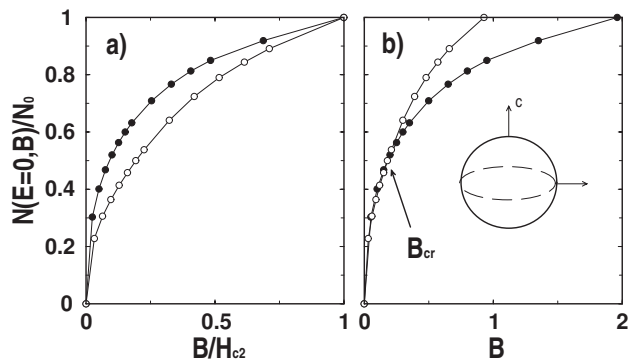}
\caption{a)
Field dependence of ZEDOS for polar state.
Scaling factor $H_{c2}$ is different for each direction.  
The best fit to low field dependence is 
$N(E=0,B)/N_0=\gamma(B)/\gamma_N\approx1.14\cdot (B/H_{c2})^{0.35}$
for $H\parallel c$ (full circles) and 
$1.06\cdot (B/H_{c2})^{0.45}$ for $H\perp c$ (empty circles).
b) ZEDOS against induction $B$
(in dimensionless units).
}
\label{machida1a}
\end{figure}

In Fig. \ref{machida1a} field dependence of  
ZEDOS for polar state  with a line node is shown for 
$H\parallel c$ (full circles)  and $H\perp c$ (empty circles).
Ratio $N(E=0,B)/N_0$ at low $T$ is equal to 
$\gamma(B)/\gamma_N$, where $\gamma(B)T$ ($\gamma_N T$) is low $T$ 
specific heat of superconducting (normal) phase.
We discuss the power-law exponent of $B$ dependence of ZEDOS. 
It is difficult to fit the data with a single power-law function 
$(B/H_{c2})^\beta$. At least in low field we can estimate 
$N(E=0,B)/N_0\sim(B/H_{c2})^{0.35}$ for $H\parallel c$, i.e. 
very steep increase with field.
Here is the explanation. The most important contribution to ZEDOS
is coming from QP that flow in the plane perpendicular to the
applied field. For $H\parallel c$ geometry, those QP 
experience zero energy gap. They are easily excited and extended 
outside of the vortex core even in low field in comparison to $s$-wave 
superconductors. The outcome is steep increase of ZEDOS with field.
Experimentally similar small exponent is observed in MgB$_2$. The
physics is analogous to the case of polar state. Small exponent is coming from 
the small gap at the $\pi$-band \cite{nakai} in MgB$_2$ while coming from the
line node in the polar state.
For perpendicular orientation $H\perp c$
(empty circles) the problem is analogous to that of 
two dimensional
(2D) d-wave case for 
fields along $c$-axis. Power law with exponent $\beta
\approx 0.45$ calculated here
should be compared with self-consistent calculation on cylindrical 
Fermi surface and 2D $d$-wave gap function which reveals 
$N(E=0,B)/N_0\sim(B/H_{c2})^{0.43}$ power law \cite{ichioka}.
For this geometry $H\perp c$, QP in plane $\perp H$
experience zero gap only if their momentum is in basal plane. Therefore,
they are more difficult to excite compared to parallel geometry $H\parallel c$,
hence the exponent $\beta$ is bigger.
 
It is important to emphasize that ZEDOS in Fig. 
\ref{machida1a}a) is plotted against
$B/H_{c2}$, where $H_{c2}$ is different for each geometry. 
For polar state there is a large anisotropy of
upper critical field $H_{c2}^\parallel\approx 2H_{c2}^\perp$.
When plotted versus magnetic field, Fig. \ref{machida1a}b), 
ZEDOS lines crosses at some critical field $B_{cr}$. 
Therefore by rotating magnetic field from $c$-axis
toward the basal plane ZEDOS may increase or decrease depending on 
field value.

\begin{figure}[h]
\includegraphics[angle=0,scale=0.45]{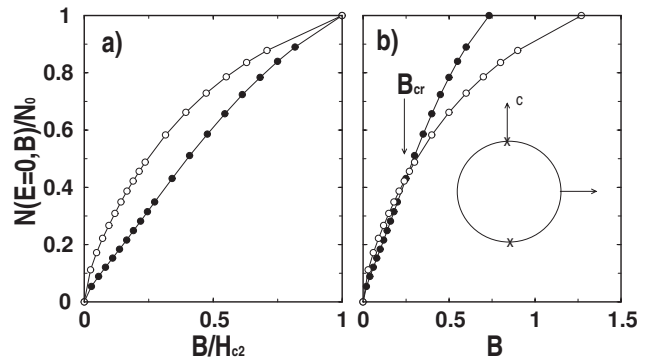}
\caption{a)
Field dependence of ZEDOS for axial state.
Scaling factor $H_{c2}$ is different for each direction. 
The best fit to low field dependence is 
$N(E=0,B)/N_0=\gamma(B)/\gamma_N\approx1.25\cdot (B/H_{c2})^{0.64}$
for $H\perp c$ (empty circles). b) ZEDOS against induction $B$
(in dimensionless units).
}
\label{machida2a}
\end{figure}

For axial state with two point nodes and $H\parallel c$, energy gap is 
small only for small
fraction of QP that flow along the $c$-axis, which makes small 
contribution to total ZEDOS. In this geometry ZEDOS resembles that in
$s$-wave superconductors. Most of the low energy QP are trapped at vortex 
cores, at least in low field, thus
$N(E=0,B)/N_0\sim B/H_{c2}$. This is confirmed by numerical 
calculation shown in Fig. \ref{machida2a}a) (full circles). 
For field $H\perp c$ (empty circles) 
power-law exponent is smaller than that in line node polar state for both
field geometries, Fig \ref{machida1a}a). 
Roughly speaking, the larger is the angular area of suppressed gap, the faster 
ZEDOS is increasing with induction $B$.
Note that the field dependence $N(E=0,B)/N_0
\sim (B/H_{c2})\ln(B/H_{c2})$ predicted by the Doppler shift 
calculation for point node case \cite{mineev} is 
far off the present result in Fig. \ref{machida2a}, warning us its validity. 
Similar to the polar state, ZEDOS curves for
two field directions cross at some critical field 
$B_{cr}$
as shown
in Fig. \ref{machida2a}b).   

\begin{figure}[t]
\includegraphics[angle=0,scale=0.4]{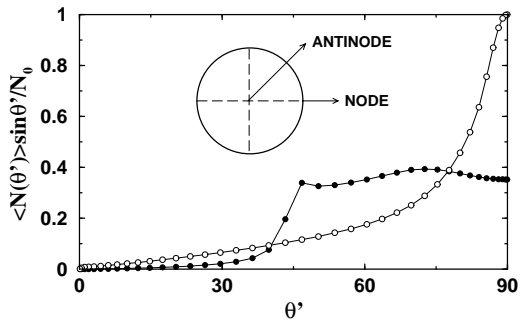}
\caption{Angle resolved ZEDOS averaged 
over angle $\phi'$ in plane perpendicular to field, 
$\left<N(\theta')\right>/N_0=(1/2\pi)
\int N(E=0,\bm v)\;d\phi'$ for antinode (full circles) and node
(empty circles) field direction. Line node is schematically presented with
dashed line in the inset. Magnetic
induction $B=0.0217H_{c2}^{node}$ ($H_{c2}^{node}/H_{c2}^{antinode}=0.828$
at $T=0.1T_c$). 
}
\label{theta}
\end{figure}

We show the importance of the mutual arrangement of line node and 
magnetic field on ZEDOS. 
In this sense it is interesting to examine 3D version 
of $k_x^2-k_y^2$ symmetry
of the gap function which is given by:
$\Omega(\phi,\theta)=\sqrt{15/4}\sin^2\theta\cos{2\phi}$.
The form of the gap function is a natural choice for
spherical Fermi surface.
The question is, 
can we guess for which field direction in the basal plane 
ZEDOS is maximum based on the calculation for polar state?
Namely,
if the magnetic field $H\perp c$ is along the node, then
all QP flowing perpendicular to the field experience zero 
energy gap (see inset in Fig. \ref{theta})
analogous to polar state with $H\parallel c$.  
If the field is along the antinode direction then 
QP that flow perpendicular to the field experience zero gap only 
if their momenta are along the c-axis, analogous to polar state and 
$H\perp c$. These simple qualitative arguments, suggesting 
$N(E=0,antinode)<N(E=0,node)$, are misleading since our calculation gives 
opposite result. It is because one must take into account the contribution 
from QP that are flowing at some angle with respect to the field
in this 3D problem.
It is instructive to see how angle resolved ZEDOS
$N(E=0,\bm v)$, averaged over 
angle $\phi'$ in plane perpendicular to field, changes with angle  
$\theta'$ between QP velocity $\bm v$ and vortex axis. We plot this quantity, 
multiplied with 
weighting factor $\sin\theta'$ in Fig. \ref{theta} at some very low field. 
Total ZEDOS  is area under the curve.
Two field directions perpendicular
to the c-axis are considered, {\em node} and {\em antinode}. For 
QP that flow perpendicular to the magnetic field, $\theta'=90^\circ$, 
ZEDOS for
node field direction is about 3 times larger than for antinode 
field direction.  However for antinode direction  QP experience gap node
as long as the angle between QP velocity and field direction is
$45^\circ<\theta'<90^\circ$.  
In spite of the  $\sin\theta'$ weighting factor there is a significant 
contribution to the total ZEDOS that are coming from these QP. 
As a result at low field $N(E=0,antinode)$ is larger than 
$N(E=0,node)$ by a few percents.

\begin{figure}[t]
\includegraphics[angle=0,scale=0.45]{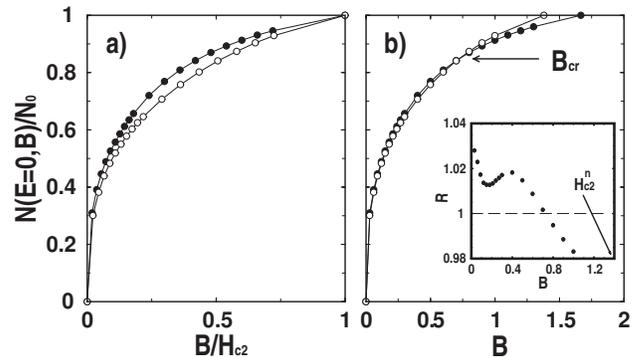}
\caption{a) Field dependence of ZEDOS for antinode (full circles)
and node (empty circles) field direction. 
The best fit to low field dependence is 
$N(E=0,B)/N_0\approx1.14
\cdot (B/H_{c2})^{0.324}$
(antinode) and $N(E=0,B)/N_0\approx1.07
\cdot (B/H_{c2})^{0.327}$ (node). b) 
ZEDOS is plotted versus induction $B$
(in dimensionless units). In the inset 
ratio $R=N(antinode)/N(node)$ is shown 
as a function of induction; $H_{c2}^n$ denotes upper
critical field along node direction.
}
\label{machida3b}
\end{figure}

In Fig. \ref{machida3b}a) field dependence of ZEDOS  
for node and antinode field direction is shown. 
Exponent $\beta\approx 0.32$ for antinode direction 
differs from $\beta\approx 0.45$  of polar state ($H\perp c$) in 
Fig. \ref{machida1a}.  This difference comes from the 
different power expansion of the gap function in the node vicinity.
In the antinode case, in plane perpendicular to the field,
gap function can be approximated as $|\Omega(\phi,\theta)|\approx
\Omega_0\theta^2$ in the vicinity of gap node $\theta=0$. On the other hand,
$|\Omega(\phi,\theta)|\sim
|\theta\pm \pi/2|$ near node $\theta=\pm\pi/2$ for polar state and $H\perp c$.
The latter case is analogous to 2D d-wave function and $H\parallel c$ giving
$\beta\approx 0.45$. 
It was shown by     
Barash and Svidzinsky \cite{barash} that temperature dependence of 
specific heat
is closely related to the exponent $n$ of gap function power
expansion near node. The larger is exponent $n$ the faster specific
heat  $C_s$ is
increasing with $T/T_c$. Similar qualitative arguments can be applied 
to field dependence of specific heat. 

At $B=B_{cr}$ two ZEDOS curves cross, the same as 
for polar and axial state, see Fig. \ref{machida3b}b).
In the inset of Fig. \ref{machida3b}b), ratio
$R=N(antinode)/N(node)$ is plotted against induction $B$. 
Physics of the crossing is very
simple and will be explained on the 3D d-wave case. In the present model 
Fermi surface is assumed to be sphere and 
upper critical field anisotropy is determined by the gap function.
For our simple 3D d-wave case $H_{c2}^{node}<H_{c2}^{antinode}$.
Therefore, for a fixed high field 
$N(E=0,node)>
N(E=0,antinode)$ because along the node field direction
the superconductor is closer to the normal state and 
ZEDOS is closer to the normal state value $N_0$.
On the other hand, for $B\ll H_{c2}$ the QP excitations probe
the gap structure since the biggest contribution to the ZEDOS is coming from
the delocalized QP. It was calculated  
$N(E=0,node)<N(E=0,antinode)$. 

The value of crossing field $B_{cr}$ depends on the Fermi surface model.  
Upper critical field is also affected by the Fermi surface 
anisotropy, and can reverse the sign of four-fold $H_{c2}$ oscillations in the
basal plane. For example, in YNi$_2$B$_2$C the gap node is along
$[100]$, $[010]$ directions, implying that those are also the directions of
$H_{c2}$ minima. 
But in borocarbides Fermi surface is highly anisotropic.
If we accept that LuNi$_2$B$_2$C has similar electronic
structure as Y-borocarbide, then minimum of upper critical field along
$[110]$ direction \cite{metlushko} implies the decisive role of Fermi surface 
on $H_{c2}$ anisotropy. Thus, in this case the high field
inequality should be 
$N(E=0,node)<N(E=0,antinode)$.
Since we expect that Fermi surface anisotropy has no role in low-field
ZEDOS, then the sign of four-fold ZEDOS oscillation should be
the same for all $B$, i.e. there is no crossing of two ZEDOS lines.

\begin{figure}[t]
\includegraphics[angle=0,scale=0.4]{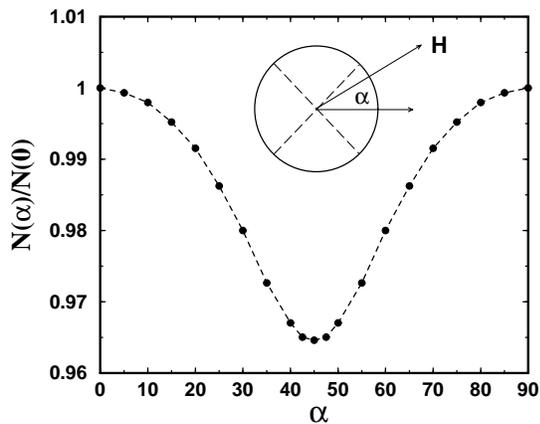}
\caption{
ZEDOS at $B=0.0217H_{c2}^{node}$ as a function of angle $\alpha$ between 
the applied field and antinode direction. 
}
\label{angular}
\end{figure}

In Fig. \ref{angular} low field angular dependence of 
ZEDOS for field rotating in the basal plane is shown. Four-fold oscillations 
is what one expects from the symmetry of the gap function. Angular variation 
is  $\approx 3\%$ at low fields, which is measurable with present 
experimental techniques \cite{park}. Note that in the 2D d-wave case 
Doppler-shift calculation \cite{vekhter} estimates angular variation as large 
as $30\%$. Parabolic-like minimum in angular dependence of ZEDOS is in 
contrast with cusp-like minimum in 2D d-wave case 
(and cylindrical Fermi surface) \cite{vekhter,hirschfeld,schachinger}.
Cusp-like minimum in thermal conductivity angular dependence is predicted
in the $s+g$ model (point node) and observed experimentally \cite{izawa}.
While shape of ZEDOS minimum can rule out some
forms of the gap function it can not provide the unique answer on the question 
of node topology (line or point). 
It is necessary to tilt the field out of the basal plane and study  
ZEDOS to gain additional information. This was done by measuring thermal 
conductivity \cite{izawa} and 
specific heat \cite{park} with fields rotating around $c$-axis
in YNi$_2$B$_2$C.

We have studied the orientational field-dependence of the nodal QP 
with zero-energy in the mixed state for the three representative gap 
functions, namely the axial, polar and d$_{x^2-y^2}$ states. Our computation 
is based on quasi-classical approach for 3D Fermi sphere. We have 
demonstrated that the orientational dependent and angle-resolved specific heat 
measurements are an ideal tool to distinguish line and point nodes and to 
locate the nodal direction free from scattering time or localization effects 
associated with transport experiments and also that this can be feasible in 
the present-day technical limitations.
When conducting field-rotation experiment, it is important to keep the field 
low ($B<B_{cr}$) to probe the intrinsic gap structure.

We thank Y. Matsuda, T. Sakakibara, K. Izawa, I. Vekhter,
M. B. Salamon and T. Park  for useful discussions.

\end{document}